\documentclass[12pt]{article}
\usepackage[pctex32]{graphicx}
\linethickness{0.4pt}

\begin{document}
\vskip 4.0 true cm

\begin{center}
{\LARGE \bf The Effects of a Non-Ferroelectric Slab on the Polarization and the Susceptibility of the Ferroelectric Multilayer}
\vskip 1.2  true cm
 Yin-Zhong Wu,$^{1,2}$ Dong-Lai Yao$^{1}$, Wen Dong$^{1}$, and Zhen-Ya Li$^{3,1}$
\vskip 0.2  true cm
$^{1}${\it Department of Physics, Suzhou University, Suzhou, 215006, China$^*$}\\
$^{2}${\it Department of physics, Changshu College, Changshu, 215500, China}\\
$^{3}${\it CCAST(World Laboratory), P. O. Box 8730, Beijing 100080, China }\\
\end{center}
\vskip 1.0 true cm

\begin{abstract}
 
The polarization and the susceptibility of a ferroelectric multilayer with a non-ferroelectric slab are investigated within the framework of transverse Ising model with a four-spin interaction term. The effect of the thickness and the position of the non-ferroelectric slab are investigated in this paper. We find that the increase of the thickness of the non-ferroelectric will decrease the polarization and the susceptibility of the film. If the position of the non-ferroelcetric slab shifts from the center of the film to the surface, the number of the peaks of the susceptibility will change. And a step-like polarization curve is found.
\vskip 0.5 true cm

\end{abstract}
\vskip 0.2 true cm
\hskip 1 true cm{\bf Keywords}: Ferroelectric heterostructure, Non-ferroelectric layer.\\

$*$ Mailing Address in China\\
E-mail: yzwu@csgz.edu.cn or zyli@suda.edu.cn
\newpage
{\bf I. INTRODUCTION}\\

During the past decade, there has been considerable interest in study of the ferroelectric heterostructure, such as ferroelectric film, multilayer, and superlattice. Heterostructure ferroelectric materials have important applications in ferroelectric memory, sensor, microelectromechanical system, and optical devices. Shen et al.[1] studied a ferroelectric sandwich structure using the Ginzburg-Landau phenomenological theory, they investigated the effects of the long-range interaction and the thickness of the middle layer on the polarization of the ferroelectric sandwich structure. Wang[2,3] and Kaneyoshi[4] applied the transverse Ising model(TIM) to investigate the phase transition properties of the ferroelectric film, ferroelectric/ferroelectric superlattice, and ferroelectric bilayer.  The ferroelectric/paraelectric superlattice was studied in experiments[5-9] and in theories[10-11]. The enhancement of remnant polarization is found in the epitaxial $BaTiO_{3}/SrTiO_{3}$ superlattice with asymmetric structure[9], which is about three times of the $BaTiO_{3}$ single-phase film formed under the same condition. And they proposed that the ferroelectric/paraelectric strained superlattice or multilayer was one of the effective approaches to enhance the remnant polarization of ferroelectric heterostructure. 
However, No study on the ferroelectric film with a non-ferroelectric slab is found in the literature. The non-ferroelectric slab inside the ferroelectric multilayer may be impurity or polymer, etc., which is inserted to the ferroelectric film to enhance the polarization and dielectric properties, or to improve the mechanical properties of the ferroelectric multilayer. Here we define the inserted slab as a non-ferroelectric slab uniformly. If a non-ferroelectric material is introduced into the ferroelectric film, what kind of interesting phenomenon, if any, will happen? What is its effect on the surface and the susceptibility of the film? This is  the motivation of our paper. 
As is well known, the long-range interactions dominate in the ferroelectric materials[12-14].  We use the transverse Ising model with long-range interaction to investigate the ferroelectric film with a non-ferroelectric slab. Since most ferroelectric heterostructure, studied in the experiments, are the ferroelectric material with first-order phase transition, four-spin interaction[15,18,19] is included in the transverse Ising model(TIM). The TIM is simple, however it is convenient. It has been successfully applied in the studying of the surface and size effects in finite ferroelectric system. We use the TIM to investigate the effects of the thickness of the ferroelectric slab and that of the non-ferroelectric slab on the average polarization, the surface polarization, and the susceptibility of the ferroelectric multilayer. We obtain that the thicker non-ferroelectric slab will lower the polarization, phase transition temperature, and the susceptibility of the film. The position of the non-ferroelectric slab will have an important role on the polarization and the susceptibility of the film. A step-like polarization is found when the non-ferroelectric slab sites on an asymmetric position. We give a detailed discussion about the origin of the step-like polarization in Sec.III.

\vskip 2.0 true cm

{\bf{II. Model and Formulation}}\\

\qquad 
We consider a sandwich structure composed of two-sided ferroelectric slabs and a middle non-ferroelectric slab with different number of layers. Each layer is defined on the $x$-$y$ plane.  And pseudo spins site on a square lattice(see Fig. 1). For ferroelectric with a first-order phase transition, a term of four-spin interaction[18,19] should be included in the transverse Ising model. The system can be described by the following Ising Hamiltonian in the presence of a transverse field,
\begin{equation}
H=-\sum_{<ij>}J_{ij}S_{i}^{z}S_{j}^{z}-\sum_{ijkl}J_{ijkl}S_{i}^{z}S_{j}^{z}S_{k}^{z}S_{l}^{z}-\sum_{i}\Omega_{i}S_{i}^{x}-2\mu E\sum_{i}S_{i}^{z},
\end{equation}
where $\Omega_{i}$ is the transverse field. $S_{i}^{z}$ and $S_{i}^{x}$ are components of spin-1/2 operator at site $i$. $\mu$ is the dipole moment on site $i$, and $E$ is the applied electric field. $J_{ij}$ and $J_{ijkl}$ are the two-spin and four-spin interaction coupling constant respectively. Since the properties of the surface and the interface are different from those of the bulk, we distinguish the interactions of the pseudo spins and the transverse field on the surface and the interface from those inside the ferroelectric slabs.  

\begin{equation}
J_{ij}=\left \{
\begin{array}{ll}
\frac{J_{s}}{r_{ij}^{\sigma}}, &{\rm for } \ i, j\ {\rm on \ the \ surface \ layer },\\\nonumber
\frac{J_{b}}{r_{ij}^{\sigma}}, &{\rm for } \ i,j   {\rm\ in \ slab \ A}, \\\nonumber
\frac{J_{in}}{r_{ij}^{\sigma}}, &{\rm for } \ i,j   {\rm\ on \ the  \ interface}, \\\nonumber
\frac{J_{b}}{r_{ij}^{\sigma}}, &{\rm for } \ i {\rm\ in \ the \ upper \ slab \ A}, j\  {\rm in \ the \ down \ slab \ A}, \nonumber
\end{array}
\right.
\end{equation}

\begin{equation}
J_{ijkl}=\left \{
\begin{array}{ll}
J_{4s}, &{\rm for } \ i, j, k, l \ {\rm on \ the \ surface },\\\nonumber
J_{4b}, &{\rm for } \ i, j, k, l   {\rm\ inside \ slab \ A}, \\\nonumber
J_{4in}, &{\rm for } \ i, j, k, l   {\rm\ on \ the \ interface }, \nonumber
\end{array}
\right.
\end{equation}

\begin{equation}
\Omega_{i}=\left \{
\begin{array}{ll}
\Omega_{s}, &{\rm  for } \ i  {\rm\ on \ the \ surface }, \\ \nonumber 
\Omega_{b}, &{\rm  for }  \ i  {\rm\ inside \ ferroelectric \ slabs}, \nonumber  \nonumber\\
\Omega_{in}, &{\rm  for }  \ i  {\rm\ on the interface }, \nonumber  \nonumber
\end{array}
\right.
\end{equation}
where  $J_{b}$ is the nearest-neighbor coupling constant inside slab $A$. $J_{s}$ is the nearest-neighbor pseudo-spin interaction constant on the surface of slab A. $J_{in}$ is the nearest neighbor pseudo-spin interaction constant on the interface.  $r_{ij}$ is the distance between lattice site $i$ and $j$ ($r_{ij}\geq 1$), $\sigma$ is introduced to describe the magnitude of the long-range interaction. 
Here, the long-range interaction is supposed to decay as $1/r^{3}$ at large distance[22], and we take $\sigma = 3$ in this paper. Considering that the environment of different site at the same layer is identical, we assume that the average value of the pseudo spin in the same layer of the ferroelectric slab has the same value. Within the frame work of the mean-field approximation and the decoupling approximation $<s_{j}^{z}s_{k}^{z}s_{l}^{z}>\approx <s_{j}^{z}><s_{k}^{z}><s_{l}^{z}>$ , the average pseudo spin along $z$ direction in $ith$ layer can be expressed as following[16]:
\begin{equation}
R_{i}=<S_{i}^{z}>=\frac{<H_{i}^{z}>}{2|H_{i}|}\tanh{\frac{|H_{i}|}{2k_{B}T}},
\end{equation}
where
\begin{equation}
<H_{i}^{z}>=\sum_{j}J_{ij}R_{j}+\sum_{jkl} J_{ijkl}R_{j}R_{k}R_{l}+2\mu E,\nonumber \\
\end{equation}

\begin{equation}
|H_{i}|=\sqrt {\Omega_{i}^{2}+(<H_{i}^{z}>)^{2} },
\end{equation}
where $i$ runs over all the sites within the ferroelectric multilayer structure.
In order to make the computation practicable, the long-range interaction is cut off at the 14th-neighbor in our calculations. 
It indicates that the value $R_{i}$ is related to $R_{j}$ for $|r_{ij}|\leq 4$ (according to our cut-off approximation). The error of the cut-off approximation can be neglected and the feasibility of the cut-off approximation has been discussed in our previous work[17]. When $i$ runs over all the ferroelectric layers in the multilayer, the above Eq. (5) forms a set of simultaneous nonlinear equations from which $R_{i}$ can be calculated numerically. Then, the polarization of the $ith$ ferroelectric layer is

\begin{equation}
P_{i}=\left \{
\begin{array}{ll}
n\mu R_{i}, &{\rm for }\ i \ {\rm \in \ the \ two-sided  \ ferroelectric \ slabs} , \nonumber \\\nonumber
0, &{\rm for }\  i \ {\rm  \in \ the \ non-ferroelectric \ slab}, \nonumber \nonumber
\end{array}
\right.
\end{equation}
where $n$ is the number of pseudo spins in a unit volume. The average polarization of the multilayer is obtained:
\begin{equation}
P_{av}=\frac{1}{N}\sum_{i=1}^{N}P_{i},
\end{equation}
where $N$ is the total layers of the multilayer structure(including the non-ferroelectric layers).\\

The dielectric susceptibility of the ferroelectric multilayer is defined as
\begin{equation}
\chi (T)=\frac{\partial P(E,T)}{\partial E}|_{E=0}=\frac{1}{N}\sum_{i=1}^{N} 2n\mu\frac{\partial R_{i}(E,T)}{\partial E}|_{E=0}.
\end{equation}
 The deviation $\frac{\partial R_{i}}{\partial E}|_{E=0}$ can be obtained by numerical differential calculation, then $\chi (T)$
is obtained numerically. By changing the position and the thickness of the non-ferroelectric slab, we investigate the effects of the non-ferroelectric slab on the polarization and the susceptibility of the ferroelectric film.\\

{\bf III. RESULTS AND DISCUSSIONS}\\
  
As reported[9], the increase of the lattice parameter $c$ is caused by the mismatch of the in-plane lattice parameters between $BaTiO_{3}$/$SrTiO_{3}$. Here we take $J_{in}>J_{b}$     
in order to ensure the enhancement of the polarization on the interface at a finite temperature. The coupling constant $J_{s}$ on the surface is weaker than the bulk in the general case. In this paper, we select $J_{in}=2.0$, $J_{s}=0.5$, and $J_{b}=1.0$. Different ratios of the four-spin and two-spin interaction coupling constant in the pure ferroelectric bulk material and ferroelectric film will result in many interesting phenomena[18,19], such as the change of the order of the phase transition. If $J_{s}/J_{b}<1$ and $J_{4s}/J_{4b}<1$, the polarization and the phase transition temperature are reduced at the surface. If $J_{in}/J_{b}>1$ and $J_{4in}/J_{4b}>1$, both of the polarization and the phase transition will increase at the interface[15]. We take the four-spin interaction coupling constant as $J_{4s}/J_{s}=J_{4b}/J_{b}=J_{4in}/J_{in}=6.0$. The selected parameters will ensure that the order of phase transition of the corresponding bulk ferroelectric material is first-order, and the polarization of the interface between the ferroelectric layer and the non-ferroelectric layer will be enhanced.\\

  Firstly, we investigate the effect of the thickness of the film on the polarization 
and the phase transition temperature of the film with fixed thickness of the middle non-ferroelectric slab. Let $L_{b}=2$, and $L_{a}=L_{a}^{'}=2, 5, 8$. From Fig. 2(a), we can see that the average polarization and the phase transition temperature of the film will increase with the increase of $L_{a}$($L_{a}^{'}$). And the surface polarization also follows this tendency (See Fig. 2(b)). If $L_{a}=L_{a}^{'}$, the ferroelectric film with a middle non-ferroelectric slab can be considered as two symmetric coupled ferroelectric film without defects. So, the increase of the thickness of the ferroelectric film with a fixed thickness of the non-ferroelcetric slab is equivalent to the increase of the thickness the ferroelectric film without defects. It is the reason why the polarization and the phase transition temperature increase with the increase of the thickness of the film with fixed value of $L_{b}$.\\
 
  From Fig. 3, we can see as the increase of the non-ferroelectric slab, the average polarization of the film will decrease, and susceptibility of the film will also decrease. There are two reasons for the decrease of the average polarization. The first reason is the increase of the component of the non-ferroelectric slab. The second reason is that the corelation between the upside slab and the underside slab becomes weak when the thickness of the non-ferroelectric slab increases. The decrease of the susceptibility is resulted by the increase of the component of the non-ferroelectric slab that is irrelevant to the external electric field.  
 
 In order to investigate the position effects of the non-ferroelectric slab, the thickness of the non-ferroelectric and the total thickness of the film are fixed as $L_{b}=1$, $L_{total}=11$. The effects of the position of the non-ferroelectric slab on the polarization of the film and the upper surface are given in Fig. 4(a) and 4(b) respectively. It is found that the polarization of the upper surface and the film decrease with 
the decrease of the distance between the non-ferroelectric layer and the upper surface 
at low temperature, and increase with the decrease of the distance at the region of 
high temperature. The phase transition temperature goes up with the decrease of the distance.     
From Fig. 4, we can see a step-like polarization, which is also found in the pure ferroelectric
thin film within the framework of the Landau Theory[20]. As the shift of the position of the non-ferroelectric slab to the upper surface, the upper surface will become disordered more easily, compared with the whole film, as the increase of the temperature. Supposing only the nearest neighbor interactions are considered, the surface phase transition temperature will be lower than the film. This step-like polarization is not corresponding to the ferroelectric-ferroelectric phase transition defined in ref.[21]. Here no phase transition is taken place at the temperature ($T_{s}$) where a step-like polarization occur. Due to the exist of the long-range interaction, the upper surface is still a ferroelectric state for $T > T_{s}$. We can safely say that the occurrence of the step-like polarization is associated with a partial disorder of the pseudo-spins at the surface. If the long-range interaction is not taken into account, the surface will become complete disordered at $T=T_{s}$. The more the non-ferroelectric slab near the surface, the larger polarization "step" will occur.      \\

  The change of the susceptibility with the change of the position of the non-ferroelectric slab is given in Fig. 4(c). When the position of the non-ferroelectric slab shifts to the upper surface, two peaks is observed in the susceptibility curve, and the two peaks correspond to the disorder of the pseudo spins in the upside slab and the underside slab of the film respectively. As the decrease of $L_{a}$, the first peak will shift to the region of low temperature, and the second peak will shift to the high temperature region. When the non-ferroelectric slab sites on the middle position of the film($L_{a}=L_{a}^{'}=5$), only one peak of the susceptibility is found. This is because the upside slab and the underside slab are symmetric when the non-ferroelectric slab sites on the center of the thin film, the two peaks will merged into one.\\

In summary, we have investigated the polarization and the susceptibility of a ferroelectric film with first-order phase transition. The effects of the non-ferroelectric slab on the film can be concluded as follows: (1) The increase of the thickness of the non-ferroelectric slab will lower the polarization of the film and the susceptibility of the film. (2)When the position of the non-ferroelectric shifts to the upper surface, a step-like polarization is found in the surface, and the phase transition temperature of the ferroelectric film will be heighten, the number of the peaks of the susceptibility will change from one to two.

\qquad 
\\       

 \vskip 1.0 true cm

{\bf ACKNOWLEDGMENTS}\\
  The work was supported by the National Science Foundation of China under the grant No.10174049.
 
\newpage
\parindent=0.0cm
$^{1}$ J. Shen and Y. Q. Ma, J. Appl. Phys. {\bf 89}, 5031(2001)\\
$^{2}$ Y. G. Wang, W. L. Zhong, and P. L. Zhang, Phys. Rev. B{\bf 55}, 11439(1996);\\
       J. M. Wesselinowa, Phys. Stat. Sol. (b) {\bf 223}, 737(2001)\\
$^{3}$ N EI Aouad, B Laaboudi, M Kerouad and M Saber, J. Phys: Condens. Matter {\bf 13}, 791(2001)\\
C. L. Wang and D. R. Tilley, Solid State Commun. {\bf 118}, 333(2001)\\       
$^{4}$ T. Kaneyoshi, Phys. Stat. Sol. (b) {\bf 220}, 951(2000);\\             
       Y. Z. Wu, D. L. Yao and Z. Y. Li, J. Appl. Phys. {\bf 91}, 1482(2002)\\
$^{5}$ H. Tabata, H. Tanaka, and T. Kawai, Appl. Phys. Lett. {\bf 65}, 1970(1994) \\
$^{6}$ E. D. Specht, H. M. Christen, D. P. Norton, M. F. Chisholm, and L. A. Boatner, Phys. Rev. Lett. {\bf 80}, 4317(1998)\\
$^{7}$ J. C. Jiang, X. Q. Pan, W. Tian, C. D. Theis, and D. G. Karkut, Appl. Phys. Lett. {\bf 74}, 2851(1999)\\
$^{8}$ E. D. Specht, H.-M. Christen, D. P. Norton, and L. A. Boatner, Phys. Rev. Lett. {\bf 80}, 4317(1998)\\
$^{9}$ T. Shimuta, O. Nakagawara, T. Makino, and S. Arai, J. Appl. Phys. {\bf 91}, 2290(2002)\\
$^{10}$ J. Zhang, Z. Yin, and M. S. Zhang, Thin Solid Films {\bf 375}, 255(2000)\\
$^{11}$ M.Sepliarsky, S. R. Phillpot, D. Wolf, M. G. Stachiotti, and R. L. Migoni, Phys. Rev. B {\bf 64}, 060101(2001)\\
$^{12}$ J. Shen and Y. Q. Ma, Phys. Rev. B {\bf 61}, 14279(2000)\\
$^{13}$ H. M. Christen, E. D. Specht, D. P. Norton, M. F. Chisholm, and L. A. Boatner, Phys. Rev. B {\bf 72}, 2535(1998)\\
$^{14}$ Y. G. Wang, W. L. Zhong and P. L. Zhang, Solid State Commun. {\bf 101}, 807(1997)\\
$^{15}$ J. M. Wesselinowa, Solid State Commun. {\bf 121}, 89(2002)\\
$^{16}$ M. G. Cottam, D. R. Tilley, and B. Zeks, J. Phys. C: Solid State Physics {\bf 17}, 1793 (1984)\\
$^{17}$ Y. Z. Wu, D. L. Yao, and Z. Y. Li, Phys. Stat. Sol. (b) {\bf 231}, 2(2002)561\\
$^{18}$ J. M. Wesselinowa, M. S. Marinov, Int. J. Mod. Phys. {\bf 6}, 1181(1992)\\
$^{19}$ C. L. Wang, Z. K. Qin, D. L. Lin, Phys. Rev. B {\bf 40}, 680(1989)\\
$^{20}$ J. F. Scott, H. M. Duiker, Paul D. Beale, et al, Physica B 150(1988)160 \\       
$^{21}$ X. S. Wang, C. L. Wang, and W. L. Zhong, Solid State Communications {\bf 122}(2002)311\\
$^{22}$J. S. Slater, Phys. Rev. 78, 748-61 (1950).
\newpage  
\begin{center}{CAPTIONS OF FIGURES}\end{center}

{\bf Fig. 1:}\\

The schematic graph of a ferroelectric multilayer structure with a non-ferroelectric slab.\\

{\bf Fig. 2:}\\

The surface polarization and the average polarization of the ferroelectric multilayer versus temperature for different thickness of the ferroelectric slab $L_{a}$ ($L_{a}^{'}=L_{a}$, $L_{b}=2$).\\

{\bf Fig. 3:}\\

The surface polarization, the average polarization, and the susceptibility of the ferroelectric multilayer versus temperature for different thickness of the non-ferroelectric slab $L_{b}$ ($L_{a}^{'}=L_{a}=5$).\\

{\bf Fig. 4:}\\

The surface polarization, the average polarization, and the susceptibility of the ferroelectric 
multilayer versus temperature for selected positions of the non-ferroelectric slab in the ferroelectric film. $L_{b}=1$, $L_{total}=11$.\\

\newpage
\vfil\includegraphics[scale=0.7]{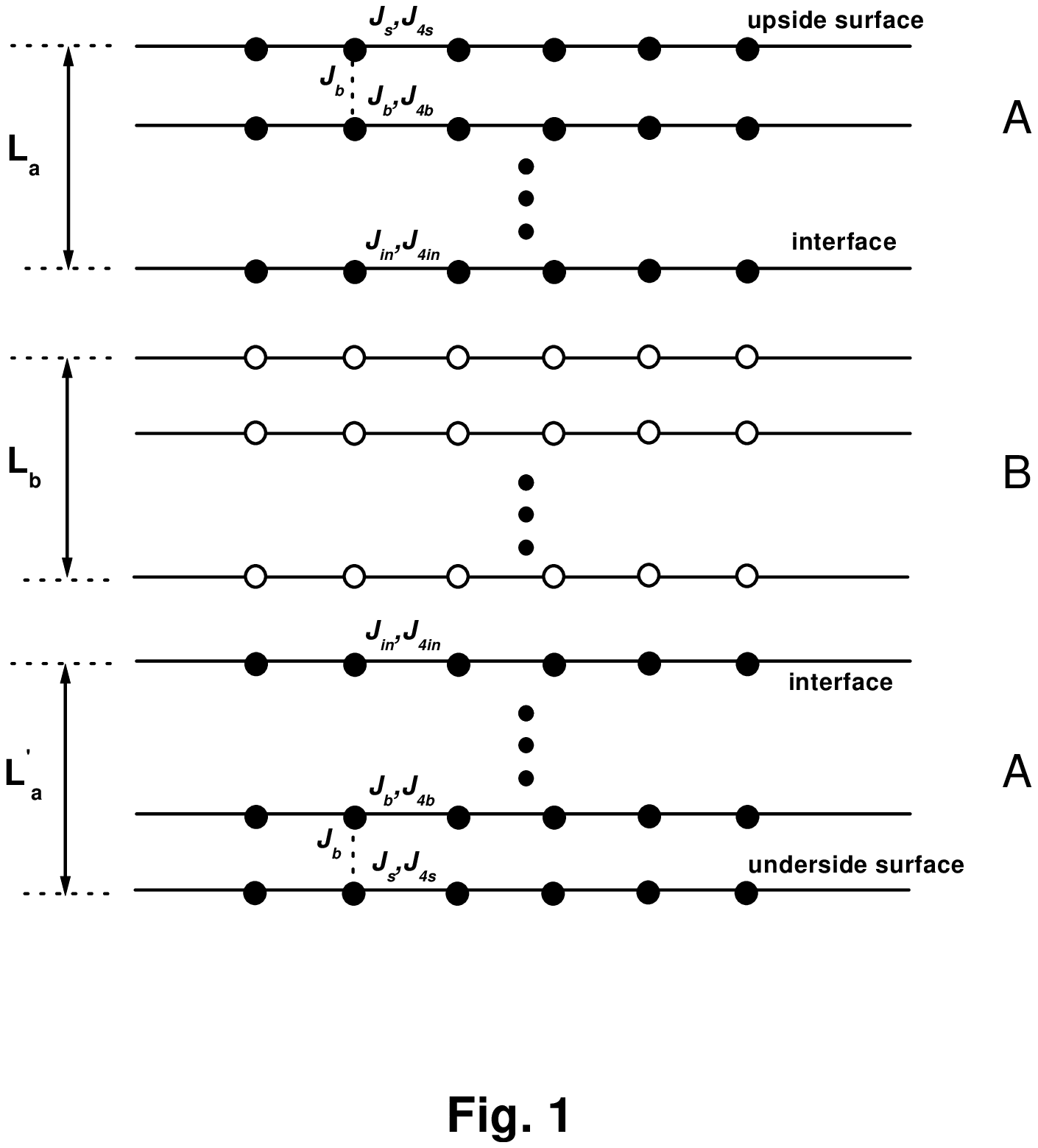}\vfil

\newpage
\vfil\includegraphics[scale=0.7]{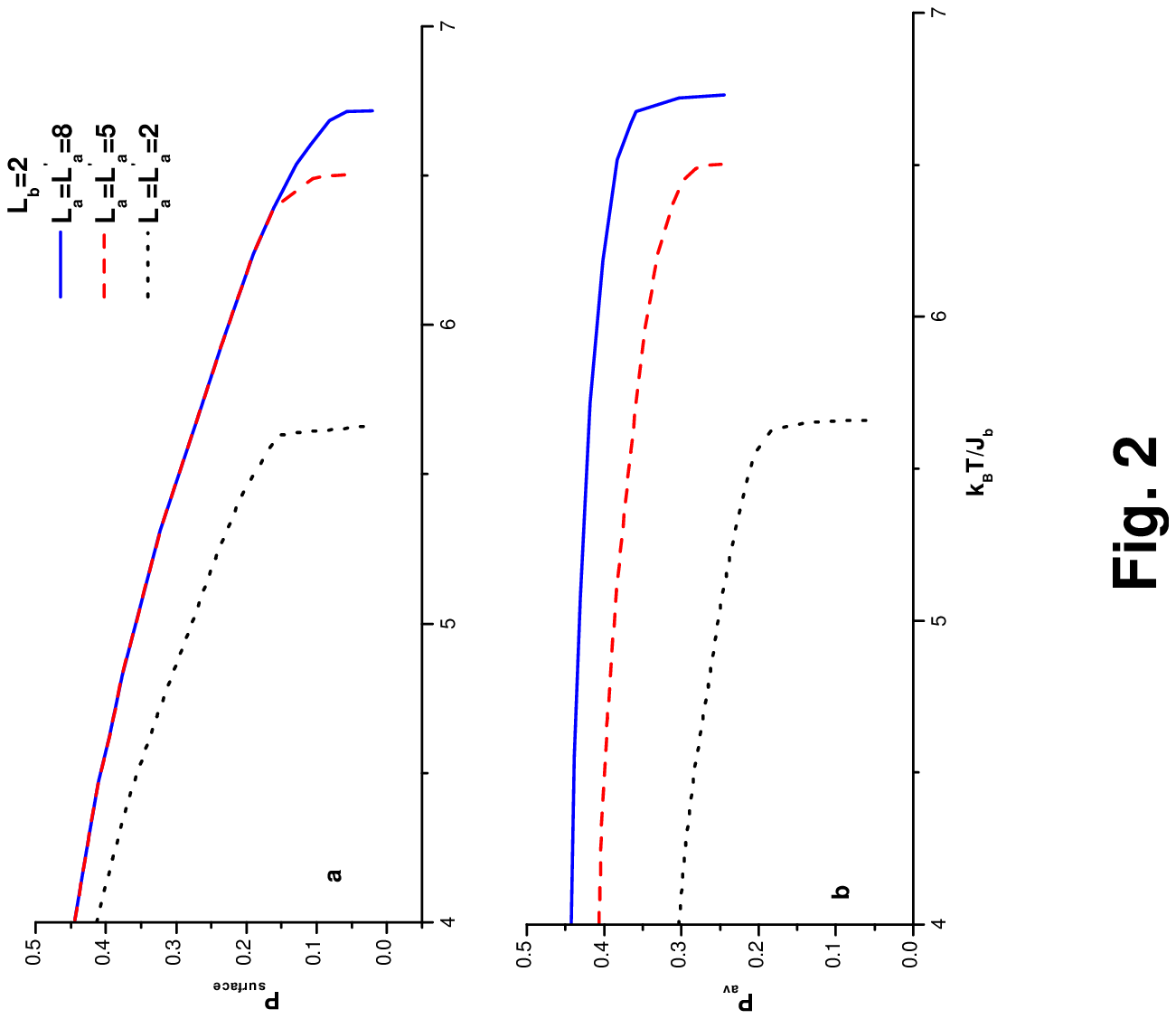}\vfil

\newpage
\vfil\includegraphics[scale=0.7]{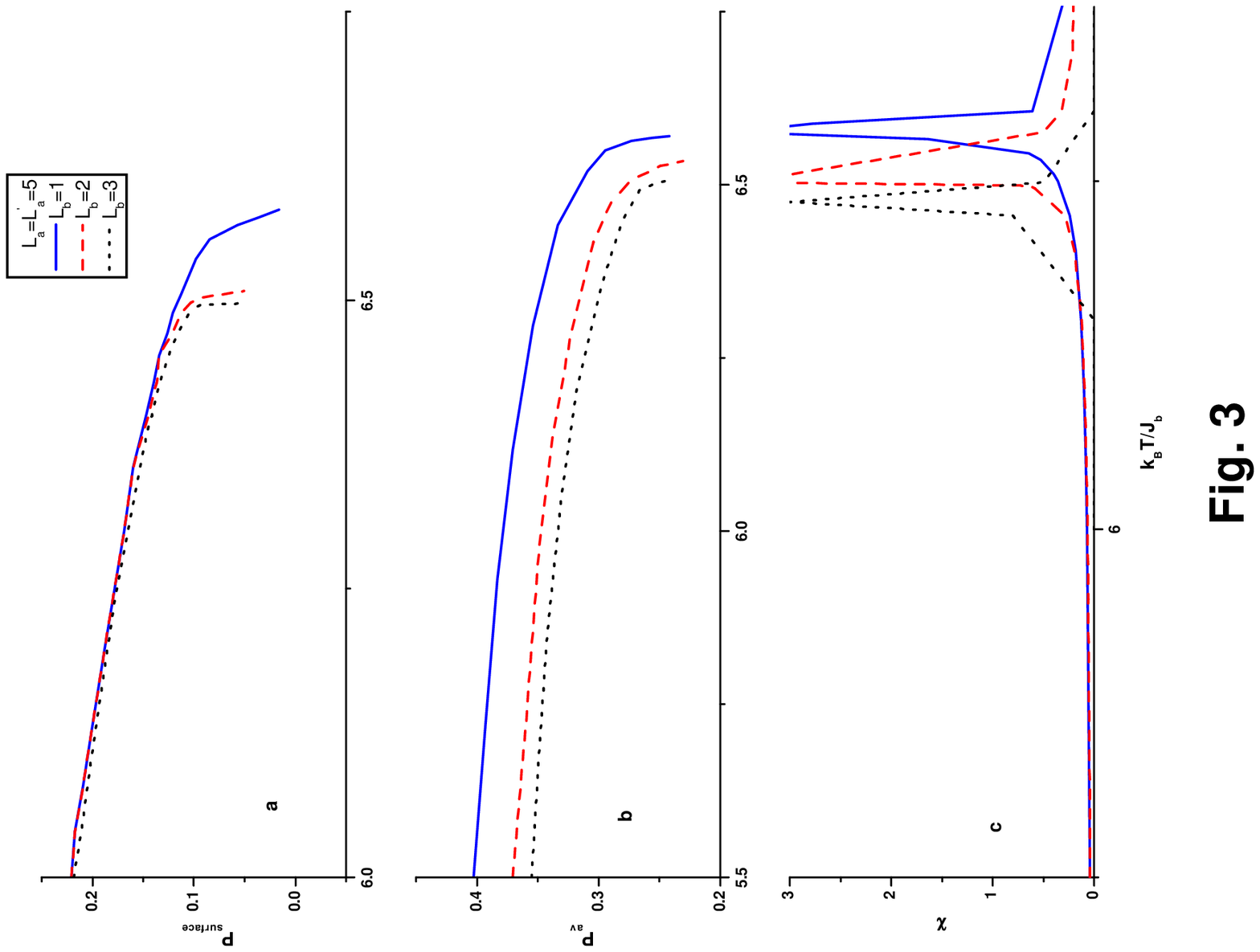}\vfil

\newpage
\vfil\includegraphics[scale=0.7]{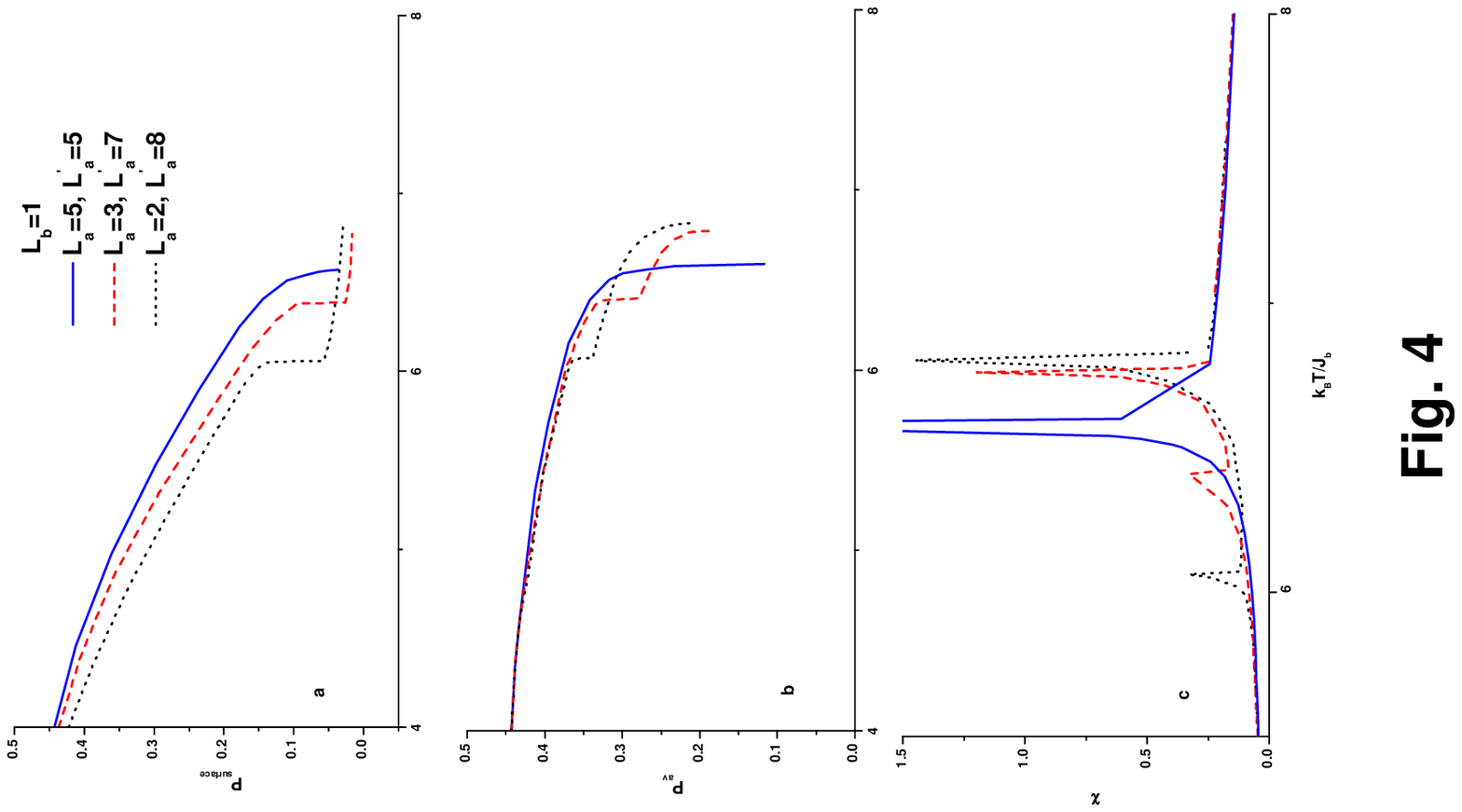}\vfil

\end{document}